\title{Computational Prediction of Muon Stopping Sites:  a Novel Take on the Unperturbed Electrostatic Potential Method}

\author{Simone Sturniolo$^1$ \and Leandro Liborio$^1$}
\date{
    $^1$Theoretical and Computational Physics Group, \\
 Scientific Computing Department, \\
 Rutherford Appleton Laboratory, STFC, \\
 Harwell Campus, Didcot, OX11 0QX.\\
    \today
}

\documentclass[ ]{article}

\usepackage{graphicx}
\usepackage{dcolumn}
\usepackage{bm}
\usepackage[mathlines]{lineno}
\usepackage[utf8]{inputenc}
\usepackage[T1]{fontenc}
\usepackage{mathptmx}
\usepackage{amssymb}
\usepackage{afterpage}
\usepackage{rotating}
\usepackage{float}
\usepackage{listings}
\usepackage{tabularx}
\usepackage{hyperref}
\usepackage{footnote}
\usepackage{longtable}
\usepackage[font=small]{caption}

\newcommand{\deftype}[2]{\textsc{Default:} \texttt{#1} \textsc{Type:} \texttt{#2}}

\begin{document}

\maketitle

\begin{abstract}
Finding the stopping site of the muon in a muon-spin relaxation experiment is one of the main problems of muon spectroscopy, and computational techniques that make use of quantum chemistry simulations can be of great help when looking for this stopping site. The most thorough approach would require the use of simulations - such as Density Functional Theory (DFT) - to test and optimise multiple possible sites, accounting for the effect that the added muon has on its surroundings. However, this can be computationally expensive and sometimes unnecessary.  Hence, in this work we present a software implementation of the Unperturbed Electrostatic Method (UEP): an approach used for finding the muon stopping site in crystalline materials that calculates the minima of the crystalline material's electrostatic potential, and estimates the muon stopping site relying on the approximation that the muon's presence does not significantly affect its surroundings. One of the main UEP's assumptions is that the muon stopping site will be one of the crystalline material's electrostatic potential minima.  In this regard, we also propose some symmetry-based considerations about the properties of this crystalline material's electrostatic potential. In particular, which sites are more likely to be its minima and why the unperturbed approximation may be sufficiently robust for them. We introduce the Python software package \texttt{pymuon-suite} and the various utilities it provides to facilitate these calculations, and finally, we demonstrate the effectiveness of the method with some chosen example systems.
\end{abstract}

\section{\label{sec:level1}Introduction:\protect}

In a muon spin spectroscopy experiment, a beam of polarized positive muons is implanted in a sample and the pattern of positron emission, caused by the muons' decay, is used to measure the magnetic properties of the host material. Positrons are emitted in the direction of the muon’s spin at the time of its decay, and the observation of the decay's time evolution can be used  to infer information about the magnetic structure of the sample.  

The muon has a half-life of $2.2\,\mathrm{\mu s}$ and, during this time, it interacts with magnetic moments of either nuclear or electronic origin. These interactions can be used to obtain information on, for instance, local magnetic fields on a small length scale.  Muons are used, therefore, mainly as a microscopic probe to measure both static and dynamic magnetic ordering. Due to the local nature of these interactions, predicting where the muon would stop inside the host material is one of the key problems of muon spin spectroscopy, and it has been the subject of many research works\cite{PhysRevB.87.115148, CAMMARERE2000636, PhysRevB.85.165211, PhysRevB.80.094524}.  

In our previous work, we presented a set of computational approaches for finding the muon stopping site that could be applied to the case of finding the stopping site of muonium, the pseudo-atom formed by a positive muon when it captures an electron. That approach requires the use of many randomly initialised calculations, based on either the Density Functional Theory (DFT) approximation\cite{doi:10.1063/1.5024450} or the much faster, lower level approximation of Density Functional Tight Binding (DFTB)\cite{doi:10.1063/1.5085197}. The DFTB approach is by far the faster of the two, but it depends on parametrisations fitted on more fundamental theory calculations that are not available for all elements of the periodic table \cite{Seifert1996,Elstner1998}. Furthermore, both approaches require a certain level of computational power, and have a learning curve associated with knowing how to properly run a DFT or DFTB simulation. However, for diamagnetic muons -namely, positive muons that have not formed a bound state with an electron and whose energetics are thus dominated by their unshielded electrostatic interaction with the surrounding environment- a much simpler alternative to find the stopping site is available: the Unperturbed Electrostatic Potential method.

The Unperturped Electrostatic Potential (UEP) method works by assuming that the muon's presence does not significantly affect the crystal order. In this assumption, one can use the known distribution of positive charges (the ions) and the negative electronic charge cloud as found, for example, via a single DFT calculation on the unperturbed system, to compute the Coulomb electrostatic potential felt by the muon at each point in the crystal. It is then possible to compute the forces acting on the muon, and which sites constitute a local or absolute minimum of the potential, without running any more expensive calculations.

The general consensus at the time of writing is that the UEP method works better in metals \cite{PhysRevB.87.115148, Bonfa2016}, where the diffuse electronic charge shields the muon effectively at very short ranges, thus preventing it from having a significant effect on the atomic positions. Conversely, if the host material is an insulator or a semiconductor, the stopping sites for the muon can be in an off-centre interstitial position because of the formation of chemical bonding between the muon and its neighbouring atoms. In this case, the unperturbed electrostatic potential isn't a satisfying methodology for finding the muon stopping site any more. This is observed, for example, in lithium fluoride, where the muon forms a F-$\mu$-F structure by distorting the lattice locally \cite{PhysRevB.87.121108}.

In addition to the UEP, we also consider the importance of symmetry considerations for determining the muon stopping sites. In general, it is considered a good rule of thumb in crystallography that sites with high symmetry are often the preferred atomic sites \cite{Souvignier2016}. This holds true for muons as well and, in practice, very often the muon stopping site will be a high-symmetry site of the crystal that has not already been occupied by another atom. For this reason, a  crystallographic analysis of the pure system that will be studied with muons is a useful first step when looking for muon stopping sites, as such analysis is significantly faster than any computer simulation. In this paper we will outline some specific considerations about the properties we can predict for the electrostatic potential at these high symmetry sites, and we will also introduce the software implementations of both the symmetry analysis and UEP in the Python library \texttt{pymuon-suite}. Finally, in Section \ref{sec:level3} we will show the results of both approaches for some choice example systems.

\section{\label{sec:level2}Theory\protect}

\subsection{Symmetry Analysis}\label{subsec:symm}

A crystal structure is defined by its symmetry properties. Any crystal, to be called such, must possess at least a translational symmetry, since it is fundamentally the infinite repetition of a finite unit cell. However, many crystals possess higher symmetry than that, having a group $\mathcal{G}$ of symmetry operations $(\mathbf{W}, \mathbf{w})$ that connect crystallographically equivalent points\cite{Souvignier2016}:

\begin{equation}\label{symm:operation}
    Y = g(X) = \mathbf{W}X+\mathbf{w} \qquad g \in \mathcal{G}
\end{equation}.

In some cases, some points $X$ can be identified for which there exists a subset $S_X \subseteq \mathcal{G}$ of operations such that $g(X) = X$ $\forall g \in S_X$; these points are called special Wyckoff positions. Wyckoff positions need not be single points; for example, the rotation axis is the Wyckoff position of a rotation operation, as it is all left unchanged by it. In this case one can say that the Wyckoff position has one free coordinate. However, we are here concerned with those Wyckoff positions that have no free coordinates, and the symmetry operations under which they are unchanged. In particular, let $X$ be a special Wyckoff position such that $g(X+\epsilon) \neq X+\epsilon$  $\forall g \in S_X$,  with $\epsilon$ an arbitrarily small vector, which is a rigorous formulation of the condition that the position has no free coordinates. It then follows that, for an arbitrary function $f(X)$ that has the same symmetry properties as the crystal, like the electrostatic potential, $X$ has to be a stationary point, namely, $\nabla f(X) = 0$.

The proof of this is the following. Consider the transformation properties of the gradient under a symmetry operation. In general, we have that $X_i' = W_{ij}Xj+w_i$, and so

\begin{equation}\label{symm:gradient}
   \nabla f = \partial_if = \partial_j'f\partial_iX_j'=\partial_j'fW_{ji} = \nabla'f\mathbf{W}
\end{equation}.

 In a special Wyckoff position, however, $\nabla f = \nabla' f$. Then it must be

\begin{equation}\label{symm:gradient2}
    \nabla f(X) = \nabla f(X) \mathbf{W}   \qquad \forall (\mathbf{W}, \mathbf{w}) \in S_X
\end{equation}.

 For Eq. \ref{symm:gradient2} to remain valid at the special Wyckoff position $X$, while the value of the gradient at that same position is different from zero, there should be at least one direction which is invariant under the entire group of operations $S_X$. But we already established that this is not the case, as the Wyckoff position X has no free coordinates. Therefore, the solution must be the trivial one, i.e.: the value of the gradient of an arbitrary function $f(X)$ at position $X$ is zero.

This already gives us a hint as to why special Wyckoff positions with no free coordinates are good candidates for the muon stopping site: because they are necessarily stationary points of the electrostatic potential. This is not even restricted only to the unperturbed system; it may still hold even if the muon introduces some deformation, provided that said deformation does not break the symmetries contained in $S_X$. However, a stationary point of a function is not guaranteed to be a minimum: it could still be a maximum or a saddle point and, in general, is not possible to determine which one it is just from symmetry considerations. However, it is possible to perform a stricter series of checks on $S_X$ to identify which Wyckoff positions put further constraints not only on the gradient, but on the Hessian of the function as well. In fact, with a similar reasoning as the one here presented for the gradient, one can find that in some points the function is constrained to have a Hessian that's either positive or negative definite (and thus, the function has either a minimum or a maximum); and in even more special cases, that the Hessian is isotropic, and the function must have locally radial symmetry. The details of this derivation are included in Appendix \ref{appendixSymmetry}.

\subsection{UEP}\label{subsec:uep-theory}

The core assumption of the Unperturbed Electrostatic Potential method is that the muon experiences a potential

\begin{equation}\label{uep_pot_dir}
    V(\mathbf{x}) = \int_{\mathbb{R}^3}
    \left[\rho_{e}(\mathbf{r})+\rho_{I}(\mathbf{r})\right]\frac{1}{|\mathbf{r}-\mathbf{x}|} d\mathbf{r}
\end{equation}

where the integral is carried out over all the (infinite) volume of the crystal, and the charge density has been split in electronic ($\rho_e$) and ionic ($\rho_I$) contributions. It can be seen easily that this integral is hard to converge in real space, as the Coulomb potential only falls off as $1/r$ whereas the Jacobian in spherical coordinates goes like $r^2$. In fact, the integral would not converge at all if the net charge wasn't guaranteed to be zero. In practice, this is always guaranteed as the system without the muon is neutral and the total charge of each unit cell is zero. However, it is far more practical to compute the integral in Fourier space, where

\begin{equation}\label{uep_pot_ft}
    V(\mathbf{x}) = \frac{4\pi}{v}\int_{\mathbb{R}^3} \left[\rho_{e}(\mathbf{G})+\rho_{I}(\mathbf{G})\right]\frac{e^{i\mathbf{G}\mathbf{x}}}{|\mathbf{G}|^2} d\mathbf{G}
\end{equation}

and convergence is faster. Here $v$ is the volume of a single unit cell, $v = (\mathbf{a} \times \mathbf{b})\mathbf{c}$ in terms of the lattice parameters.\newline
The current implementation of the UEP method in \texttt{pymuon-suite} is compatible only with CASTEP calculations, though support for other codes will be developed in the future. CASTEP is a DFT code using a plane wave basis set and pseudopotentials. The plane wave basis means it's most suited to treat periodic solids, and here we can make use of that peculiarity by noting that the electronic density returned by it can be expressed as a truncated Fourier series, and so the electronic contribution to the potential $V_e$ can be written as

\begin{equation}\label{uep_pot_elec}
    V_e(\mathbf{x}) = \frac{4\pi}{v}\sum_{\mathbf{G} > 0} \rho_{e}(\mathbf{G})\frac{e^{i\mathbf{G}\mathbf{x}}}{|\mathbf{G}|^2} 
\end{equation}

, where the sum runs over a finite number of reciprocal vectors $\mathbf{G}$ that depends on the cut off used for the calculation. Ideally, when preparing the calculation, one should have taken care that this value is high enough to properly converge the energy of the system, which means the truncation should not affect the final value much as $\rho_e(\mathbf{G})$ will go to zero for large $\mathbf{G}$. Note that we ignore $\mathbf{G} = 0$, where eq. \ref{uep_pot_elec} would diverge, because this term represents the total charge, and while we know that independently it will not be zero for both $\rho_e$ and $\rho_I$, we also know it will eventually cancel out when we sum them as that represents the fact that the system is overall neutral. \newline
The problem remains then of how to deal with $\rho_I$. This is the charge density including both the positive charge $Z_i$ of each atomic nucleus and the negative one of its $N_i$ core electrons that CASTEP embeds in the pseudopotential. We choose here to simplify the problem by assuming that this charge can be treated as a purely Gaussian charge distribution, centred on the atomic position and of width $\sigma_i$ that we base off some measure of the expected radius of the ion. Currently the software uses $\sigma_i = r_{ppot}/s$ where $r_{ppot}$ is the smallest core radius used in the construction of the pseudopotential, and $s$ is a user-defined scaling parameter. This leads to the expression

\begin{equation}\label{uep_pot_ion}
    V_I(\mathbf{x}) = \frac{4\pi}{v}\sum_i(Z_i-N_i)\sum_{\mathbf{G} > 0}e^{-\frac{1}{2}\sigma_i^2G^2} \frac{e^{i\mathbf{G}(\mathbf{x}-\mathbf{x}_i)}}{|\mathbf{G}|^2} 
\end{equation}

carrying a sum over the ions indexed with $i$, at positions $\mathbf{x}_i$.\newline
The use of this Gaussian approximation requires a little discussion. First, it should be noted that the use of the sum over only the discrete wave vectors $\mathbf{G}$ instead of an integral is perfectly legitimate: it corresponds to the fact that we're not representing only a single ion with each term, but rather, an infinite periodic array of them arranged in a crystal lattice. This means that all terms of $\rho_I$ that do not lie on points of the reciprocal lattice go to zero. We can also note that obviously the resulting charge distribution does not likely resemble the real one. However, this charge distribution is positive, as $Z_i > N_i$ for all ions, and thus repulsive to the muons. Since we are interested only in the minima of the potential for the muon, we can be sure they will be as far as possible from these lumps of positive charge! In fact, the details of the shape would matter much more for the potential at short range - from afar, these will act mostly like point charges, as their size is still rather small compared to the overall volume of the cell.\newline
Finally, we consider the problem of the cutoff over reciprocal space vectors. While for $\rho_e$ this is exact, as the finite grid is the one that CASTEP itself used to compute it in the first place, for $\rho_I$ it is effectively a truncation of an infinite series. The crucial parameter becomes then $\sigma_i$. If this is big enough, then the Gaussian term in eq. \ref{uep_pot_ion} falls off before the cutoff is even reached. If it's smaller, then there's a risk of the truncation having a sensible effect on the charge density and thus on the potential, producing spurious undulatory behaviour. At the moment, it is up to the user to diagnose and counteract these problems. The parameter $s$ that controls the scale of $\sigma_i$ can be adjusted to fix any issues depending on the specifics of the system of interest; in general we found that a default value of $s = 5$ tends to work well in most cases.

\section{Software Implementation}\label{sec_softimpl}

Both the symmetry analysis and the UEP implementation described here can be found within a Python library we deployed specifically to aid muon computational science, \texttt{pymuon-suite}. The library can be found on Github \cite{pymuonsuite} and is released under a GNU v3.0 open source license. The library depends on a few other libraries to work properly. The most important ones are Numpy \cite{numpy}, Scipy \cite{oliphant2007python}, the Atomic Simulation Environment \cite{ase-paper}, Spglib \cite{spglib} and Soprano \cite{soprano}. All of these libraries are available on the Python Package Index and thus can be installed automatically together with \texttt{pymuon-suite} without any additional effort on the user's part.\newline
Once installed, \texttt{pymuon-suite} provides the user both with a Python API to use for custom programs and with a series of pre-packaged scripts that perform the most common operations. Four of these scripts are relevant to the tools described in this work:

\begin{itemize}
    \item \texttt{pm-symmetry} performs the analysis of special Wyckoff positions described in Section \ref{subsec:symm};
    \item \texttt{pm-uep-opt} performs an optimisation of the muon position under the Unperturbed Electrostatic Potential;
    \item \texttt{pm-uep-plot} produces 1D and 2D ASCII files describing the UEP potential along paths or on specified planes, useful to produce plots;
    \item \texttt{pm-muairss} produces batches of structure files with a muon defect added following a Poisson random distribution, and analyses and clusters the results of their optimisation.
\end{itemize}

The symmetry script is particularly simple, as it does not need any parameters, and it can be runs simply by executing the command

\begin{verbatim}
    pm-symmetry <structure file>
\end{verbatim}

in which the structure file has to be any supported crystallographic file format (such as \texttt{.cif} or \texttt{.cell}). For the UEP scripts, instead, input files in the YAML format \cite{yaml} containing parameters are necessary. These scripts run with

\begin{verbatim}
    pm-uep-plot <parameters file>
    pm-uep-opt <parameters file>
\end{verbatim}

. Finally, \texttt{pm-muairss} operates similarly, but has both a `write' and a `read' mode, one to create structure files, the other to interpret the results of their optimisation. It also needs both a structure and a parameter file as inputs. To avoid mistakenly overwriting important data, the default mode is read. The two modes are used as follows:

\begin{verbatim}
    pm-muairss -t w <structure file> <parameters file>      # To write
    pm-muairss -t r <structure file> <parameters file>      # To read
\end{verbatim}

. The \texttt{-t r} argument is optional. For parameters files, each script has its own arguments that can be set using them. The specific accepted parameters and their meaning for each script are detailed in Appendix \ref{appendixInput}.

\section{\label{sec:level3}Example Systems\protect}

Table (\ref{tab:results}) compiles the muon stopping sites predicted by experiments and the UEP method for all the materials studied in this work. We chose examples of different nature: copper, which is a metal; rutile TiO$_{2}$, a semiconductor;  MnSi manganese silicide, a material that exhibit a homochiral spin spiral structure; $\mathrm{Fe_{3}O_{4}\;magnetite}$, a ferrimagnetic oxide, and LiF, an insulator. In all of these materials, the muon stopping site has been unequivocally determined by experimental methods.  Table (\ref{tab:tech_details_UEP}) shows the main UEP parameters used for determining the muon stopping site for all the examples simulated in this work.

In copper, the muon stopping site was experimentally determined to be at the centre of a copper octahedron in the copper fcc crystalline lattice. Level-crossing measurements were performed at the TRIUMF muon source (Canada) for samples at temperatures of 40K and 156K and longitudinal magnetic fields ranging from 0 to 0.012T\cite{PhysRevB.43.3284}. The UEP method predicted two stopping sites in fcc copper: one in a tetrahedral site and the other in an octahedral site.  The cluster with the largest number of structures and the lowest average energy is the one that places the muon in the octahedral site.

\begin{figure}[htb]
(T)\includegraphics[scale=0.2]{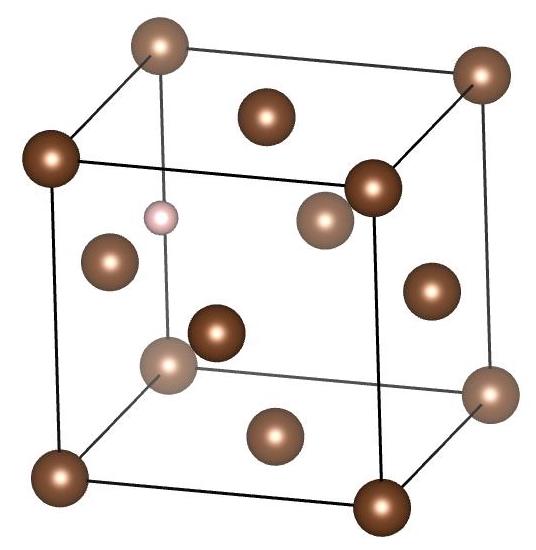} 
(O)\includegraphics[scale=0.2]{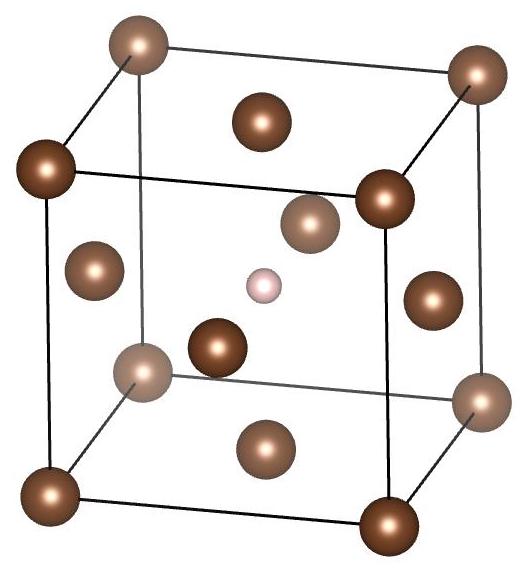} 
\caption{\label{fig:Cu} Predicted (a) tetragonal (T), and (b) octahedral (O) $\mu$ stopping sites in $\mathrm{Cu\;bcc}$. }
\end{figure}

In $\mathrm{TiO_{2}\;rutile}$, transverse Field $\mu SR$ measurements performed in the MUSR instrument at ISIS(UK)\cite{PhysRevB.92.081202} identified muon stopping sites where the muon has a low temperature ground state and a high temperature excited state: both corresponding to a muon bound to one of the six O atoms that form an octahedra around the $\mathrm{Ti^{3+}}$ at the centre of the $\mathrm{TiO_{2}\;rutile}$ unit cell. Each one of these stopping sites has a different O-$\mathrm{Ti^{3+}}$ bonding configuration, with the ground state formed by bonding the muon to the in-plane oxygens that lie in the same  plane as $\mathrm{Ti^{3+}}$.  These two sites are related by symmetry and are only distinguished by the electronic structure of the $\mathrm{TiO_{2}\;rutile}$.  The cluster with the largest number of structures and the lowest average energy predicted by the UEP method is the one that places the muon at the ground state described above. 

\begin{figure}[hb]
\includegraphics[scale=0.2]{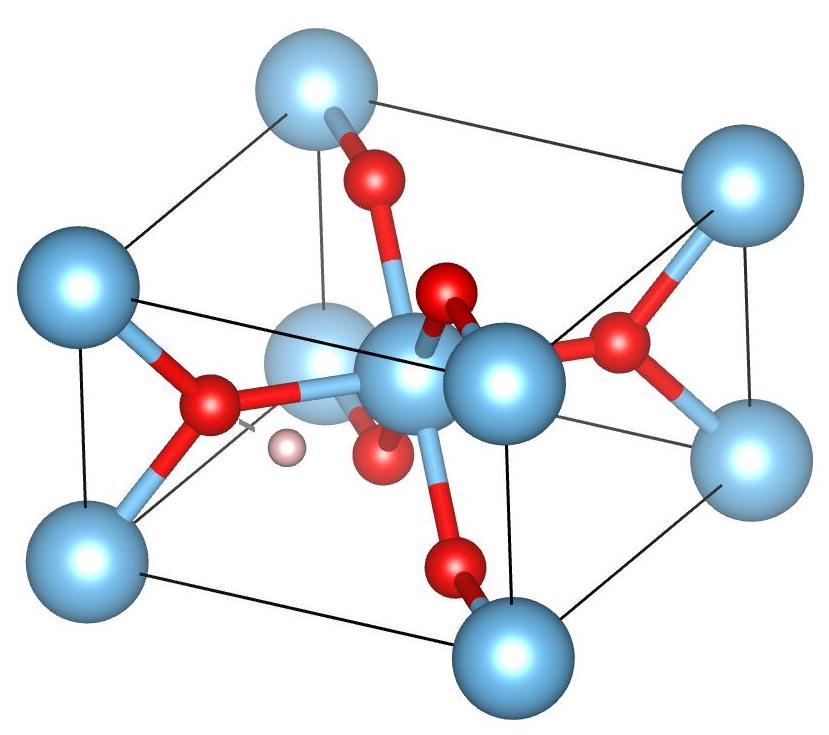}
\caption{\label{fig:rutile} Predicted $\mu$ stopping site in $\mathrm{TiO_{2}\; rutile}$.  $\mu$ close to $\mathrm{O_{apical}}$ with OH line in the ab plane.}
\end{figure}

Regarding MnSi, transverse field $\mu SR$ experiments carried out at the GPS instrument in PSI (Switzerland)\cite{PhysRevB.89.184425} identified the stopping site of the muon to be along the 4a-I Wyckoff axis of symmetry, in the MnSi unit cell.  The stopping site was identified to have the fractional coordinates given by (0.532,0.532,0.532). The UEP method predicted four potential stopping sites: a highly symmetric one, (S) and three others: A$_{1}$, A$_{2}$ and A$_{3}$. The S site, which originates from the cluster with the largest number of structures predicted by the UEP, agrees with the experimentally observed site. 

\begin{figure}[htb]
(S)\includegraphics[scale=0.12]{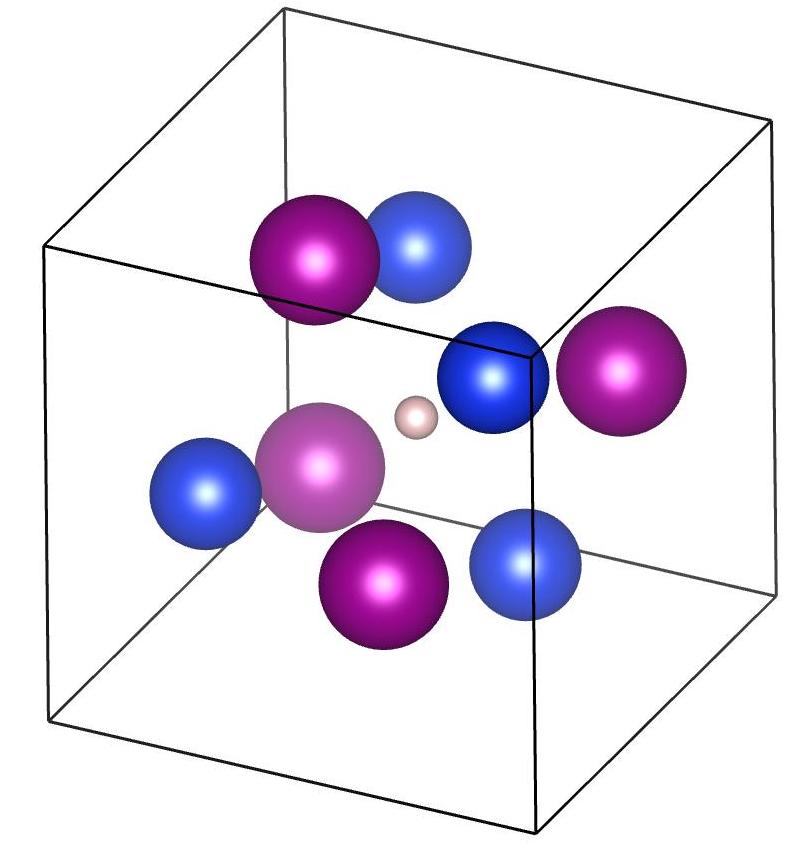} 
(A$_{1}$)\includegraphics[scale=0.12]{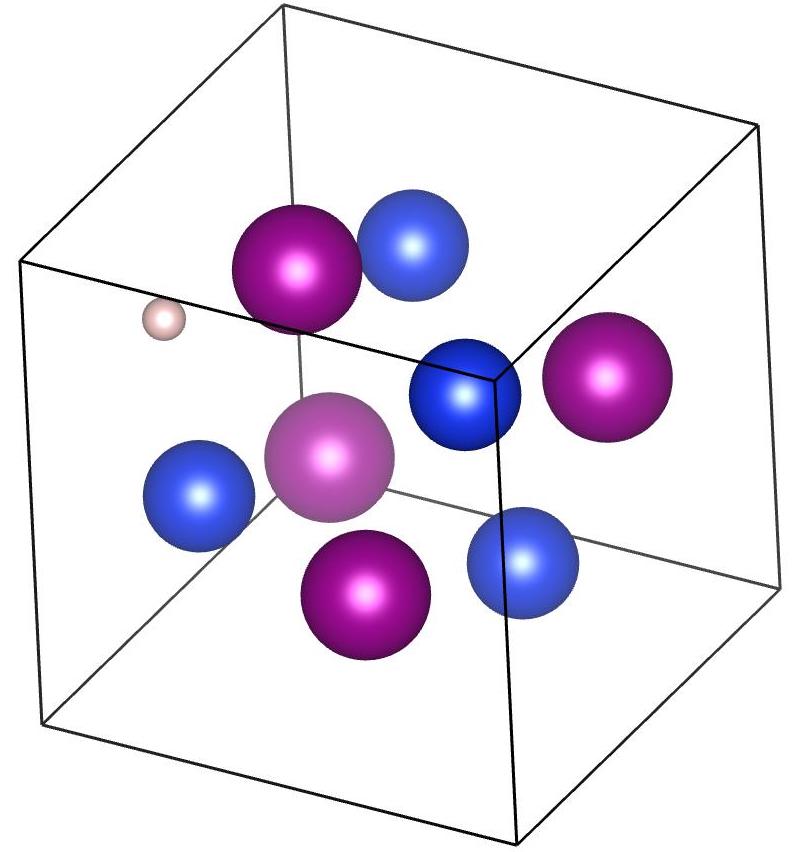} 
(A$_{2}$)\includegraphics[scale=0.12]{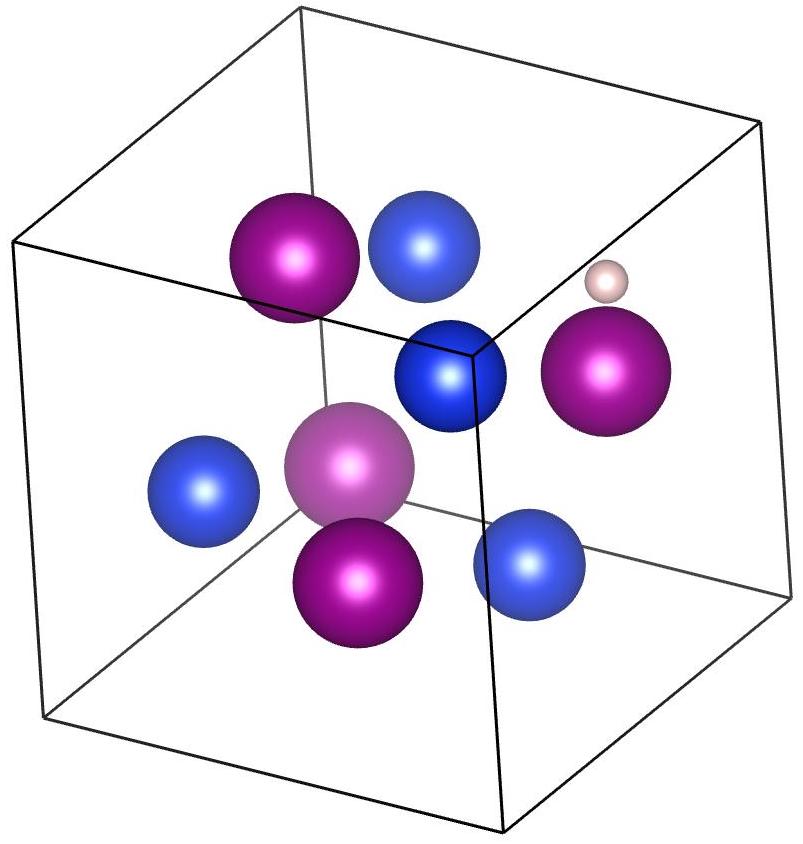}
(A$_{3}$)\includegraphics[scale=0.12]{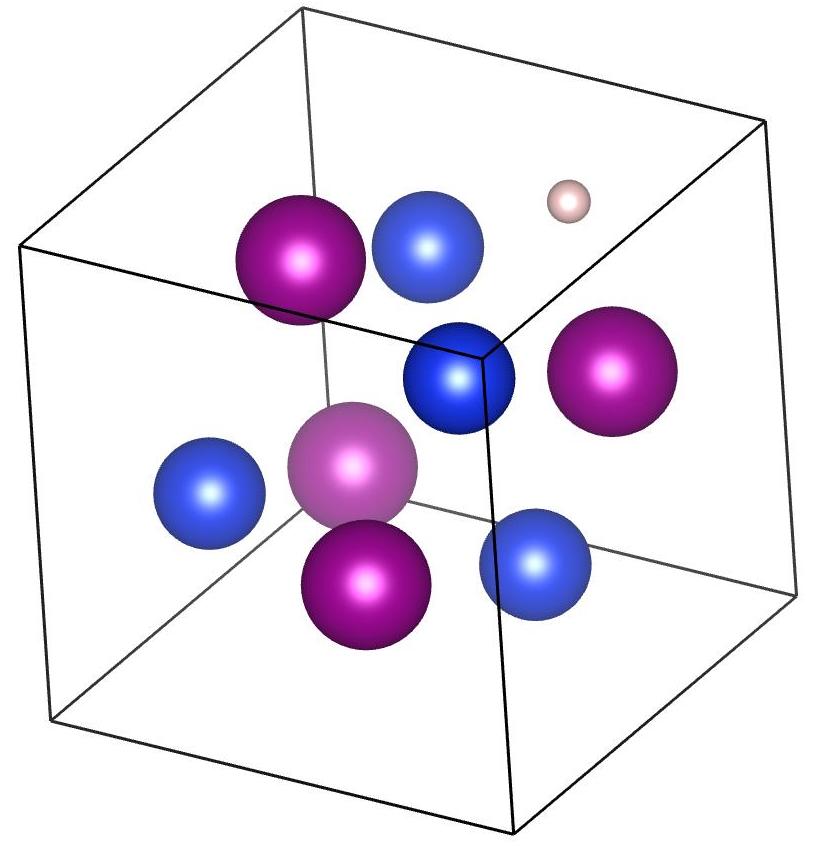}
\caption{\label{fig:MnSi} Predicted S, A$_{1}$, A$_{2}$ and A$_{3}$ $\mu$ stopping sites in $\mathrm{MnSi}$. }
\end{figure}

\begin{figure}[htb]
\includegraphics[scale=0.15]{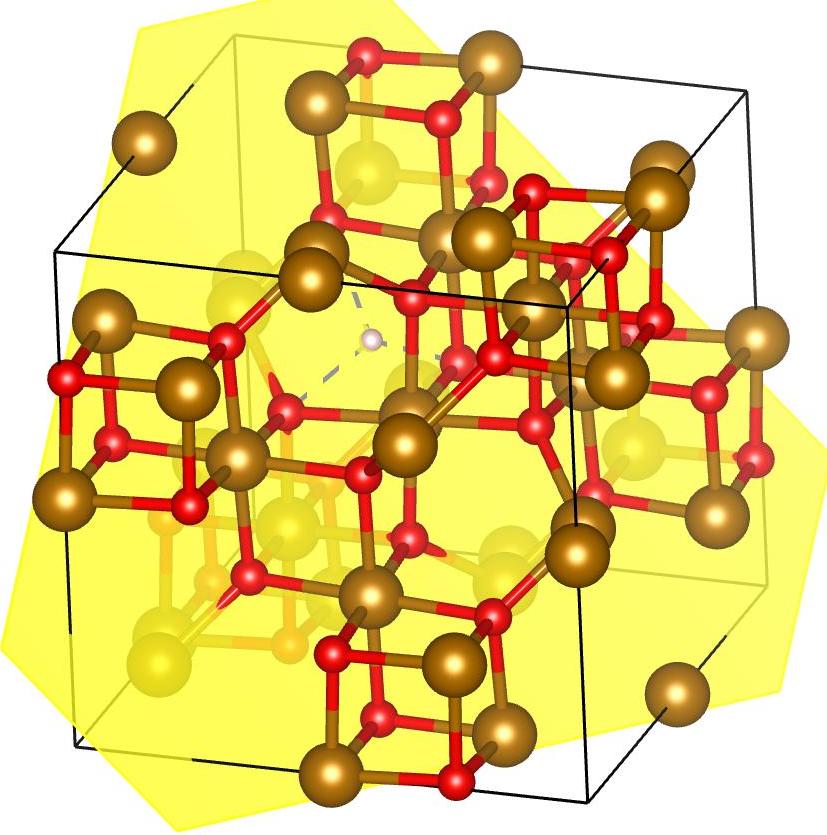} 
\includegraphics[scale=0.15]{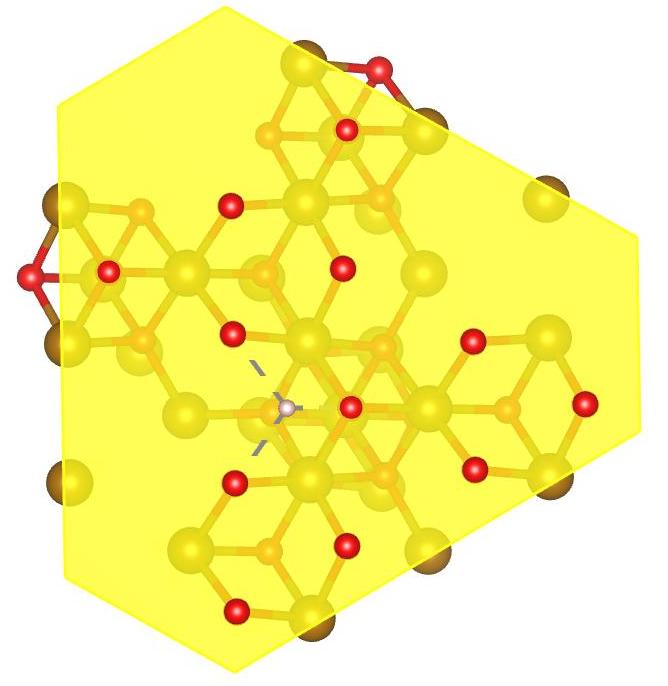} 
\caption{\label{fig:fe304-plane} Planar region for the predicted $\mathrm{O_{\langle 111 \rangle}}$ $\mu$ stopping site in $\mathrm{Fe_{3}O_{4}}$. }
\end{figure}

Potential muon stopping sites in $\mathrm{Fe_{3}O_{4}\;magnetite}$ were found using Transverse Field $\mu SR$, experiments performed at LAMPF (US)\cite{Hyper1}.  The stopping sites are: (a) located within in a planar region that is perpendicular to the  $\langle 111\rangle$ direction and, (b) situated within $\approx 1.5$\r{A} of one of the oxygen atoms defining the planar region.  The Figures (\ref{fig:fe304-plane}) show the  $\mathrm{O_{\langle 111\rangle}}$ site predicted by the UEP method. An example of a planar region perpendicular to the $\langle 111\rangle$ direction is indicated in yellow.  The muon is located at $\approx 1.3$\r{A} from its closest oxygen atom in the planar region.  

\begin{figure}[htb]
(C)\includegraphics[scale=0.15]{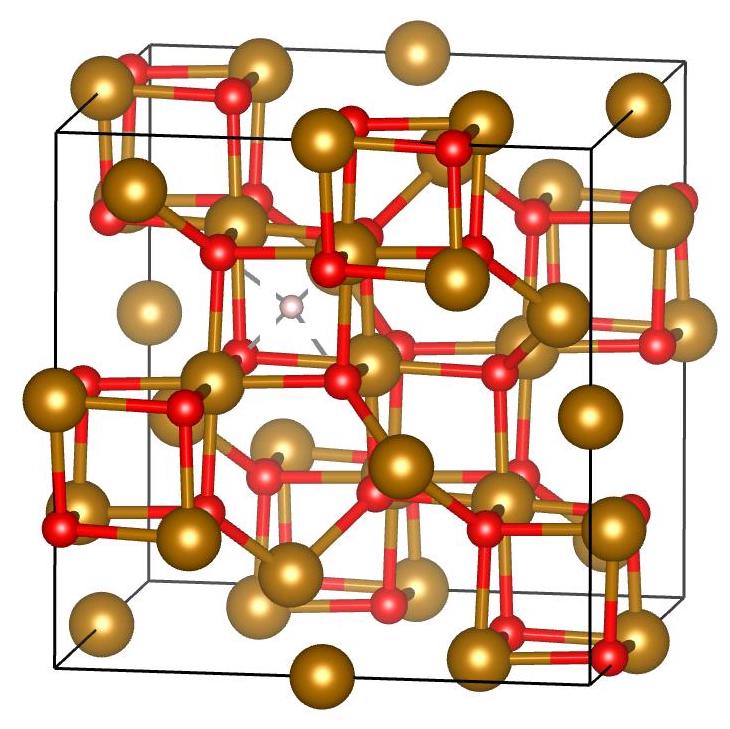} 
$\mathrm{O_{[111]}}$\includegraphics[scale=0.15]{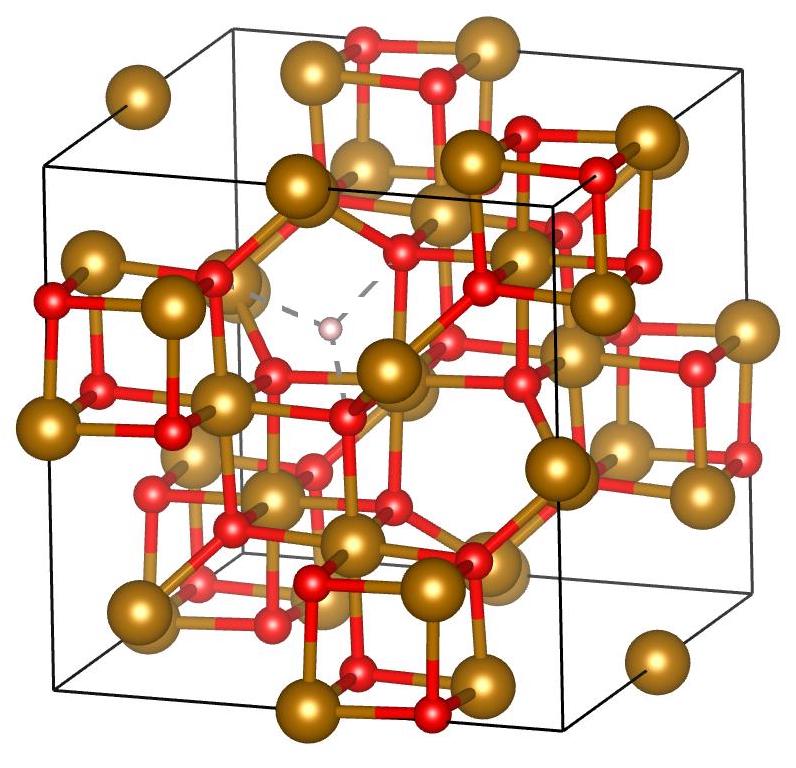} 
\caption{\label{fig:fe304} Predicted $\mathrm{O_{[111]}}$, and cubic (C) $\mu$ stopping sites in $\mathrm{Fe_{3}O_{4}}$. }
\end{figure}

\begin{figure}[htb]
(T)\includegraphics[scale=0.2]{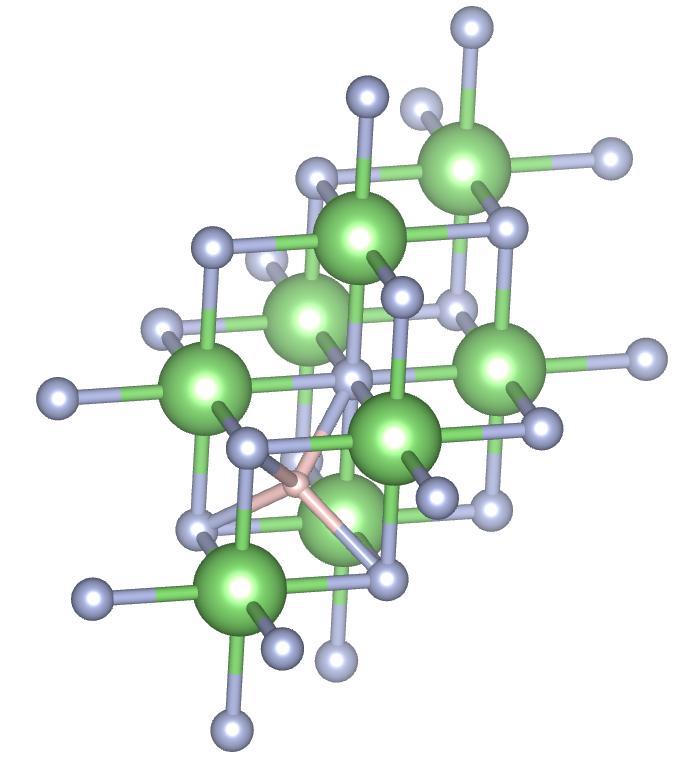} 
(V)\includegraphics[scale=0.2]{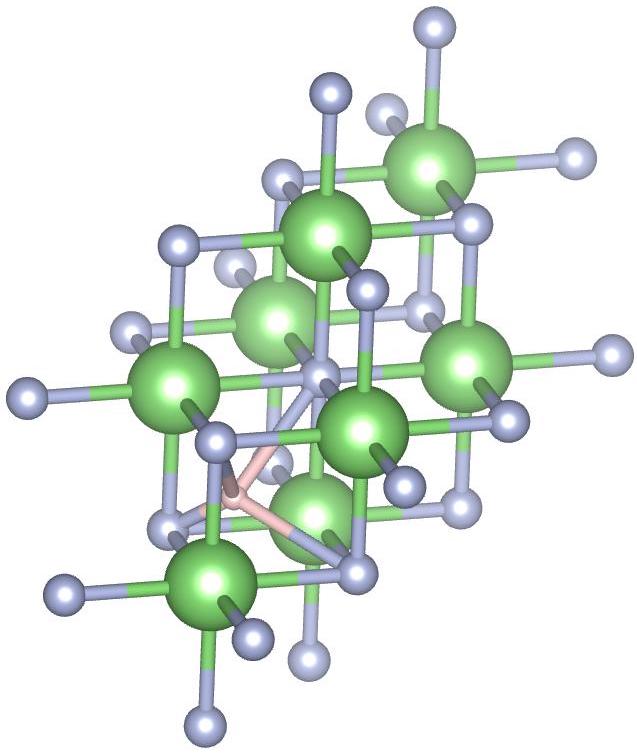} 
\caption{\label{fig:LiF} Predicted (T) tetragonal , and vertex (V) $\mu$ stopping sites in LiF.}
\end{figure}

Finally, the behaviour of $\mu$ in LiF has been studied using Zero Field $\mu SR$ experiments performed at the M15 muon channel at TRIUMF in Canada\cite{PhysRevB.33.7813}.  The stopping site is located between two F atoms forming the distinctive F$\mu$F centre. None of the predicted stopping sites are, however, in the F$\mu$F centre: they are contained within a tetrahedron defined by four F atoms, and are shown in Figures (\ref{fig:LiF}).  Site (T) is at the centre of the tetrahedron, while site (V) is displaced from the tetrahedron's centre and closer to one of its vertices. 

For all these example systems, we also performed a symmetry analysis of all their Wickoff positions using the \texttt{pm-symmetry} library, which are shown in Table \ref{tab:pm-symmetry}. 

\begin{savenotes}
\begin{table*}[htb]
\begin{tabular}{|c|c|c|}
\hline
 Sample & UEP & Wickoff positions \\ \hline
  $\mathrm{Cu\;fcc}$& O and T & 40 positions including O and T  \\
 $\mathrm{TiO_{2}\;rutile}$& $\mathrm{O_{apical}}$ and $\mathrm{O_{planar}}$ & 10 positions including $\mathrm{O_{apical}}$ and $\mathrm{O_{planar}}$ \\
  $\mathrm{MnSi}$ & S, $A_{1}$, $A_{2}$, $A_{3}$ & no empty Wickoff position$^{\textrm{*}}$. \\
  $\mathrm{Fe_{3}O_{4}\; mag.}$ & $\mathrm{O_{\langle 111\rangle}}$ and C  & 24 positions including C \\
  $\mathrm{LiF}$ & T and V & 8 positions including T \\ \hline
\end{tabular}
\caption{\label{tab:pm-symmetry} Symmetry analysis results for all the samples simulated in this work.\\
(*)The \texttt{pm-symmetry} software only identifies Wyckoff positions with \textit{zero} degrees of freedom, i.e.: points. MnSi has no Wyckoff points: it has the 4a-I Wyckoff axis.} 
\end{table*}
\end{savenotes}

\begin{figure}[h]
\includegraphics[scale=0.28]{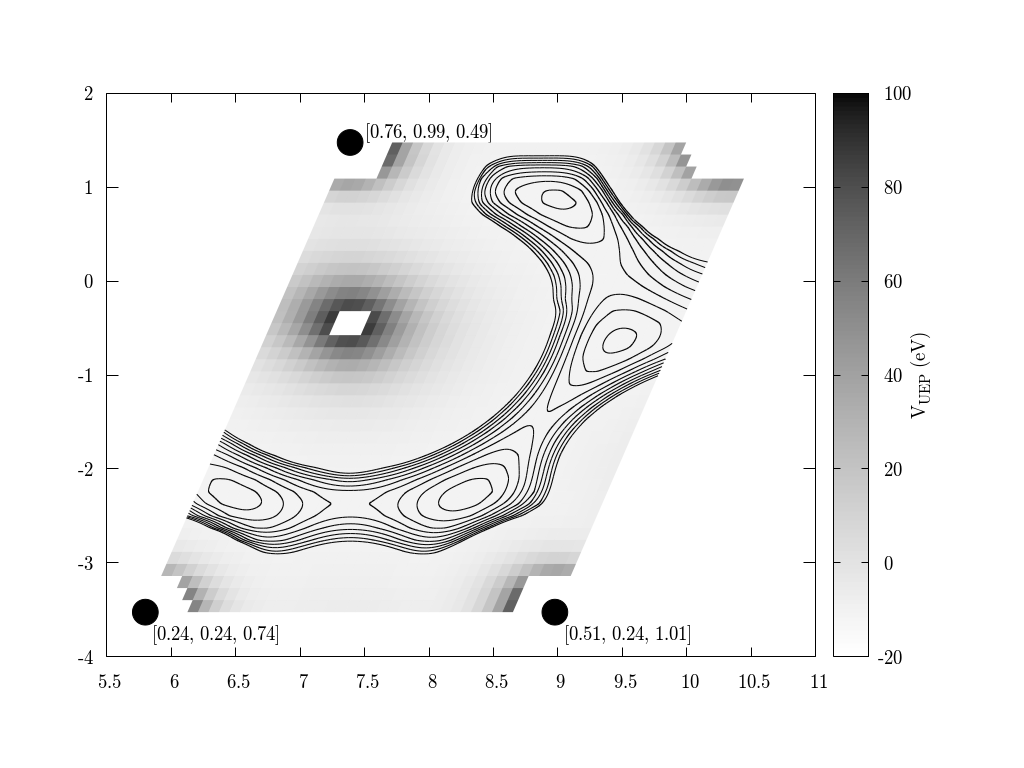}
\caption{\label{fig:fe304-UEPplane} 
UEP potential in the planar region  perpendicular to the  $\langle 111\rangle$ in $\mathrm{Fe_{3}O_{4}}$. The black dots indicate the three closest oxygen atoms to the predicted muon stopping site.}
\end{figure}

As we can see, many of the muon stopping sites theoretically predicted by the UEP method can be hinted at by performing a very simple calculation using the \texttt{pm-symmetry} library. Similarly, we can run the \texttt{pm-uep-plot} library and obtain graphic files that describe the UEP potential in a specified region of the unit cell.  An example of these UEP potential plots is shown in Figure (\ref{fig:fe304-UEPplane}), which corresponds to the planar region in $\mathrm{Fe_{3}O_{4}}$ that is perpendicular to the  $\langle 111\rangle$ direction and and is indicated in yellow in Figures (\ref{fig:fe304-plane}).

In the Supplementary Information there are examples of input and output files for  the \texttt{pm-symmetry},  \texttt{pm-muairss} and \texttt{pm-uep-opt} libraries that correspond to the search for the muon stopping sites in  $\mathrm{Fe_{3}O_{4}}$.  These examples are organised so as to explain, with a practical example, how the search of the muon stopping sites could be carried out.     

\begin{savenotes}
\begin{table}[htb]

\begin{tabular}{|c|c|c|c|} 
 \hline
 Sample & Experimental Sites & Theoretical Sites (UEP)  & Clusters \\ \hline

 $\mathrm{Cu\;fcc}$ &  octahedral$^{*}$  & octahedral (O) and  tetrahedral (T) & $\mathrm{O(\frac{67}{103}})$, $\mathrm{T(\frac{36}{103}})$ \\
 
 \hline
 $\mathrm{TiO_{2}\;rutile}$ & close to $\mathrm{O_{apical}}$ and $\mathrm{O_{planar}}$$^{**}$  & close to $\mathrm{O_{apical}}$& $\mathrm{O_{apical}(\frac{559}{563}})$  \\
\hline
  
$\mathrm{MnSi}$&  4a-I Wyckoff$^{***}$  & S, A$_{1}$, A$_{2}$, A$_{3}$  & $\mathrm{S(\frac{230}{423}})$, $\mathrm{A_{1}(\frac{104}{423}})$, \\ &&&$\mathrm{A_{2}(\frac{67}{423}})$, $\mathrm{A_{3}(\frac{22}{423}})$ \\
\hline
 
$\mathrm{Fe_{3}O_{4}\; magnetite}$ & close to O in planes $\perp$ to $\langle 111\rangle$$^{\dag}$
& $\mathrm{O_{\langle 111\rangle}}$ and cubic (C) & $\mathrm{O_{\langle 111\rangle}(\frac{139}{152}})$, $\mathrm{C(\frac{13}{152}})$ \\
\hline
 
$\mathrm{LiF}$ & F$\mu$F state$^{(\ddag)}$ & tetrahedral site ($\mathrm{T}$) and & $\mathrm{T(\frac{1}{42}})$, $\mathrm{V(\frac{41}{42}})$ \\
&& vertex site (V) & \\
\hline 

\end{tabular}

\caption{\label{tab:results} Experimental and theoretical determination of the $\mu^+$ stopping site in the samples studied in this work. The cluster column describes the fraction of muonated structures in the clusters representing the muon stopping site.  The predicted site that agrees with the experimental result is the one with the largest fraction of structures. \\
\small{
$^{(*})$Level-crossing measurements, M15 muon channel at TRIUMF (Canada).T= 40K and 156K, B=0.012T\cite{PhysRevB.43.3284}\\
$^{(**)}$Transverse Field $\mu SR$, MUSR instrument at ISIS (UK). T=[1.2-10K], B=0.02T\cite{PhysRevB.92.081202}\\
$^{(***)}$Transverse Field $\mu SR$, GPS instrument at PSI (Switzerland). T=50K, B=0.52T\cite{PhysRevB.89.184425} \\
$^{(\dag)}$Transverse Field $\mu SR$, muon channel at LAMPF (US).T= 298K, B=0.4T\cite{Hyper1} \\
$^{(\ddag)}$Zero Field $\mu SR$, M15 muon channel at TRIUMF (Canada).T= 80K\cite{PhysRevB.33.7813}
}}
\end{table}
\end{savenotes}

\begin{table*}[htb]
\begin{tabular}{|c|cccc|}
\hline 
 Sample & $\mathrm{poisson}$[\AA] & $\mathrm{vdw\_scale}$ & $\mathrm{uep\_gw\_factor}$ &$\mathrm{clustering\_hier\_t}$\\ \hline
  $\mathrm{Cu\;fcc}$& 0.4 & 0.25 & 4.0 & 0.2 \\
 $\mathrm{TiO_{2}\;rutile}$& 0.4 & 0.25 & 5.0 & 0.2 \\
  $\mathrm{MnSi}$ & 0.5 & 0.3 & 5.0 & 0.2 \\
  $\mathrm{Fe_{3}O_{4}\; mag.}$ & 0.8 & 0.5 & 6.0 & 0.2 \\
  $\mathrm{LiF}$ & 0.6 & 0.25 & 4.0 & 0.2 \\ \hline
\end{tabular}
\caption{\label{tab:tech_details_UEP} Unperturbed Electrostatic Potential technical details for all the samples simulated in this work.}
\end{table*}

\section{\label{sec:level4}Conclusions\protect}

In this work, we have presented, described and tested our new Python software package \texttt{pymuon-suite} and its various associated utilities, which can be used to study the potential muon stopping sites in crystalline materials.  We have also shown that there is a connection between the symmetry properties of the crystalline material's electrostatic potential and the potential muon stopping sites in that material.  Finally, the examples studied in this work show that, our version of the UEP method, can reliably predict the muon stopping sites in a variety of materials using purely theoretical means and reasonably cheap computer simulations. 

\section{Data Availability Statement}

The data that supports the findings of this study are available within the article and its supplementary material.

\section{Acknowledgements}

  The authors are grateful for the computational support provided by: (a) STFC Scientific Computing Department's SCARF cluster  and (b) the UK Materials and Molecular Modelling Hub for computational resources, which is partially funded by EPSRC (EP/P020194/1). Funding for this work was provided by STFC-ISIS muon source, the Ada Lovelace Centre and the CCP for NMR Crystallography, funded by EPSRC grants EP/J010510/1 and EP/M022501/1.

\newpage

\appendix
\section{\label{appendixSymmetry} Properties of the Hessian at special Wyckoff positions}

Let us consider the Hessian $H$ of a generic function with space group symmetry $\mathcal{G}$ at a special Wyckoff position $X$, invariant under all symmetry operations $g \in S_X$.  From similar considerations as in Equation \ref{symm:gradient} we can find that in general the Hessian transforms as

\begin{equation}\label{appSymm:hessianTransf}
    H = \mathbf{W}^T H' \mathbf{W}
\end{equation}
 
. Again, at $X$, it must be that $H = H'$ for all the operations under which $X$ is fixed. We know that the Hessian must be a symmetric matrix. We can then split it in an isotropic part and a traceless symmetric tensor:

\begin{equation}\label{appSymm:hessianDecomp}
    H = h_{iso}\mathbf{W}^T\mathbf{W} + \mathbf{W}^T H^{(symm)} \mathbf{W} = 
    h_{iso}\mathbb{I} + \mathbf{W}^T H^{(symm)} \mathbf{W} \qquad \forall (\mathbf{W}, \mathbf{w}) \in S_X
\end{equation}

with $h_{iso} = \mathrm{Tr}(H)/3$, and where we made use of the fact that if $X$ is a special Wyckoff position, $S_X$ can only contain reflections, rotations, inversions and rotoinversions, all of which have orthogonal transformation matrices (in general, this is not necessarily a given). This leaves us with the following relation:

\begin{equation}\label{appSymm:hessianSymm}
    H^{(symm)} = \mathbf{W}^T H^{(symm)} \mathbf{W} \qquad \forall (\mathbf{W}, \mathbf{w}) \in S_X
\end{equation}

. We can recast this problem in a way that's very similar to Eq. \ref{symm:gradient2}. This is fundamentally a system of nine equations in nine variables:

\begin{equation}\label{appSymm:hessianSys}
    H^{(symm)}_{ij} = W_{il}W_{jk}H^{(symm)}_{kl}
\end{equation}

. However, since we know that $H^{(symm)}$ is both symmetric and traceless, it only has five independent components we care about, and four of the equations are just linear combinations of the other. We can then write this as a system of five equations in five variables in matrix form:

\begin{equation}\label{appSymm:hessianFiveEq}
    (\mathbf{W}^{(5)}-\mathbb{I})H^{(5)} = 0 \qquad \forall (\mathbf{W}, \mathbf{w}) \in S_X
\end{equation}

where $H^{(5)}$ is a column vector containing the five independent components in any order we like, and the precise form of $\mathbf{W}^{(5)}$ depends on the convention we chose. One can then check for each special Wyckoff position whether there is any vector that satisfies all these conditions. The principle is the same as in Eq. \ref{symm:gradient2} - we seek whether there is at least one axis that is invariant under all the $\mathbf{W}^{(5)}$. If there is none, then $H^{(5)}$ must be necessarily zero, which means that the Hessian will be isotropic at X, $H(X) = h_{iso}(X)\mathbb{I}$.

One here must pay attention. All of these considerations apply to $X$ expressed in fractional coordinates, the ones for which the operations in $\mathcal{G}$ are written. The Hessian, too, includes the derivatives of the function with respect to the fractional coordinates. In the case of a space group with a cubic lattice, all considerations immediately apply to the Hessian in cartesian coordinates too, and one only has to divide $h_{iso}$ by the square of the lattice parameter $a$. However, in general, if we have a set of lattice vectors $\mathbf{C}$ such that the Cartesian coordinates of the Wyckoff position are $X_C = \mathbf{C}X$, then the isotropic Hessian transforms as

\begin{equation}
    H_C = (\mathbf{C}^{-1})^T H \mathbf{C}^{-1} = h_{iso}(\mathbf{C}^{-1})^T\mathbf{C}^{-1}
\end{equation}

. It then depends on the properties of the matrix $(\mathbf{C}^{-1})^T\mathbf{C}^{-1}$ what the Hessian is like in Cartesian coordinates. If the cell is cubic, then it will remain isotropic, as said above. If it is orthorombic, then it will not be isotropic, but still be positive or negative definite. In general, if $(\mathbf{C}^{-1})^T\mathbf{C}^{-1}$ is definite, then so will be the Hessian. Conversely, if it is not, we can then guarantee that the Hessian will not be either - which would make the special Wyckoff position a saddle point of the function.

\section{\label{appendixTechDetails} Technical Details of Calculations}

The DFT-based computer simulations carried out in this work were performed performed with the CASTEP\cite{Segall_2002} code.  The  plane wave cutoff, $\mathrm{E_{cutoff}}$, for these calculations was chosen by converging energy and forces. This was done using the automated tool CASTEPconv\cite{doi:10.1063/1.5085197}, to try a range of possible values, with every other condition fixed. The final choices were the values for which any successive refinement yielded a difference in energy and forces lower than a fixed tolerance. This was taken to be smaller than the tolerances used for self-consistent field and geometry optimisation calculations.  As regards the k-point grid size, Monkhorst-Pack\cite{PhysRevB.13.5188} k-point grids were used. This produced forces accurate well within an error of 0.05 eV/\AA, which was used as the limit tolerance for geometry optimisation.  

Geometry optimisation on these structures were performed with a LBFGS algorithm, with fixed unit cell parameters, to a tolerance of 0.05 eV/\r{A} for the forces. The PBE exchange-correlation functional was used, in combination with auto-generated ultrasoft pseudopotentials.

Table (\ref{tab:tech_details}) shows the main parameters used in the CASTEP calculations performed in this work. 

\begin{table*}[h]
\begin{tabular}{|c|ccc|}
\hline
 Sample &$\mathrm{E_{cutoff}[eV]}$&$\mathrm{kpoint \; grid}$ & XC Funct. \\ \hline
  $\mathrm{Cu\;fcc}$& 500.0 & 4$\times$4$\times$4 & PBE \\
 $\mathrm{TiO_{2}\;rutile}$& 700.0 & 2$\times$2$\times$2 & PBE \\
  $\mathrm{MnSi}$& 700.0 & 3$\times$3$\times$3 & PBE \\
  $\mathrm{Fe_{3}O_{4}\; mag.}$& 800.0 & 3$\times$3$\times$3 & PBE \\
  $\mathrm{LiF}$& 700.0 & 3$\times$3$\times$3 & PBE \\ \hline
\end{tabular}
\caption{\label{tab:tech_details} CASTEP technical details for all the samples simulated in this work.}
\end{table*}

\section{Parameters used \texttt{.yaml} input files} \label{appendixInput}

Here we list the keywords for the input files of the \texttt{pymuon-suite} scripts used in this work. The scripts are the ones listed in section \ref{sec_softimpl}. The keywords are then used in a file written in the YAML format, namely, a plain text file with rows written in the format

\begin{lstlisting}
<keyword>:  <value>
\end{lstlisting}



\subsection{Unperturbed Electrostatic Potential optimisation: \texttt{pm-uep-opt}} \label{app:uep_opt_keywords}

Here are listed all the keywords used to run the \texttt{pm-uep-opt} script, used for optimization of the muon position under the Unperturbed Electrostatic Potential approximation.

\begin{itemize}
	\item \texttt{chden\_path}: Path of the folder in which the charge density file produced by a CASTEP calculation (extension \texttt{.den\_fmt}) to use for the UEP can be found.\deftype{.}{string}
	
	\item \texttt{chden\_seed}: Seedname of the charge density file produced by a CASTEP calculation (extension \texttt{.den\_fmt}) to use for the UEP. In combination with the previous keyword, the file will be searched as \texttt{<chden\_path>/<chden\_seed>.den\_fmt}.
	\deftype{NONE}{string}
	
	\item \texttt{gw\_factor}:  Gaussian width factor used to define the size of the ionic charges by scaling the pseudopotential radius. Corresponds to the $s$ factor as described in section \ref{subsec:uep-theory}. \deftype{5.0}{float}
	
	\item \texttt{mu\_pos}: Starting position of the muon in the unit cell, expressed in absolute coordinates, in \AA. \deftype{[0.0, 0.0, 0.0]}{[float]}
	
	\item \texttt{geom\_steps}: Maximum number of geometry optimisation steps.
	\deftype{30}{int} 
	
	\item \texttt{opt\_tol}: Force tolerance for each geometry optimisation in eV/\AA.\deftype{1E-5}{float}
	
	\item \texttt{opt\_method}: Method used for geometry optimisation. Corresponds to one of the methods used by Scipy's \texttt{scipy.optimize.minimize} function. See the documentation for the options.\deftype{trust-exact}{string}
	
	\item \texttt{particle\_mass}: Mass of the particle, in kg. Important for zero point energy estimation. By default is the mass of the muon.
	\deftype{1.884E-28}{float}
	
	\item \texttt{save\_pickle}: If True, save the output result in a pickled file for further reading and reusing with other Python scripts.
	\deftype{true}{boolean}
\end{itemize}

\subsection{Unperturbed Electrostatic Potential plotting: \texttt{pm-uep-plot}}

Here are listed all the keywords used to run the \texttt{pm-uep-plot} script, used for plotting the Unperturbed Electrostatic Potential along directions and planes in the unit cell. We omit explaining the keywords \texttt{chden\_path}, \texttt{chden\_seed} and \texttt{gw\_factor}, which are in common with \texttt{pm-uep-opt} and work as explained in section 1 of this appendix.

\begin{itemize}
	\item \texttt{line\_plots}: Specify one or more line segments along which to plot the value of the UEP. Each line segment is specified by a list, and there are a number of possible methods to specify them:
	\begin{itemize}
		\item crystallographic direction, starting point, length and number of points. For example
		
		\begin{verbatim}
			-	[[1, 1, 0], [0, 0, 0], 10, 100]
		\end{verbatim}
		
		will produce a plot along the [110] direction, starting from the origin, continuing for 10 \AA{} and with 100 points spaced 0.1\AA{} each.
		
		\item starting point, end point, number of points. For example
		
		\begin{verbatim}
			- [[0, 0, 0], [1, 1, 1], 100]
		\end{verbatim}
		
		will produce a plot sampling the vector connecting the origin with the position [1, 1, 1] \AA{} in the cell (these are absolute positions), with a grid of 100 points.
	 
	 \item starting atom, end atom, number of points. For example
	 
	 \begin{verbatim}
		 -	[1, 2, 20]
	 \end{verbatim}
	 will produce a plot sampling the line connecting the atoms with indices 1 and 2 in the structure file, split in 20 points.		
	\end{itemize}
	
	\item \texttt{plane\_plots}:  Specify one or more planes along which to plot the value of the UEP. Each plane is specified by a list, and there are a number of possible methods to specify them:
	\begin{itemize}
		\item three corners, points along width, points along height. For example
		
		\begin{verbatim}
			-	[[0, 0, 0], [3, 0, 0], [0, 0, 3], 20, 20]
		\end{verbatim}
		
		in a cubic lattice with $a=3$ would produce a plot of the $xz$ face of the cell, split into a 20x20 grid.
		\item three atom indices to act as corners, points along width, points along height. For example	
		\begin{verbatim}
		-	[0, 1, 2, 20, 20]
		\end{verbatim}
		
		would produce a parallelogram having the vector connecting atoms 0 and 1 as base, the vector connecting atoms 0 and 2 as side, and 20 points along each side, for a 20x20 overall grid.
	\end{itemize}
\end{itemize}

\subsection{Random structure generation: \texttt{pm-muairss}}

Here are listed all the keywords used to run the \texttt{pm-muairss} script. For completeness, words that are relevant to usage with CASTEP or DFTB+ for structure optimisation are included too, though they are not relevant for the current work.

\begin{itemize}
	\item \texttt{name:} Name to call the folder for containing each structure. This name will be postfixed with a unique number, e.g. \texttt{struct\_001}. \deftype{struct}{string}
	
	\item \texttt{calculator:} Calculator(s) used to optimise the muon position. Must be a comma seperated list of values. Currently supported calculators are CASTEP, DFTB+, and UEP. Can also pass ALL as an option to generate files for all calculators. 
	\deftype{dftb+}{string}
	
	\item \texttt{poisson\_r:} Radius in \AA{} for generating muon sites with the Poisson disk algorithm.  This radius is the minimum distance at which two muons can be placed from each other when the muonated structures are generated.  \deftype{0.8}{float}
	
	\item \texttt{uep\_chden:} CASTEP charge density file.  \texttt{seed.den\_fmt}.  \deftype{NONE}{string}
	
	\item \texttt{uep\_gw\_factor:}  Gaussian width factor used to define the size of the ionic charges by scaling the pseudopotential radius. Corresponds to the $s$ factor as described in section \ref{subsec:uep-theory}. \deftype{5.0}{float}
	
	\item \texttt{vdw\_scale:} Scale factor to multiply the standard Van der Waals radius of each atom in the system, used to determine the minimum distance allowed between a muon and other atoms. Bigger values will evacuate a larger sphere around the existing atoms.\deftype{0.5}{float}
	
	\item \texttt{charged:} Determines whether the implanted muons will be charged or neutral. Must be True to use UEP. \deftype{false}{boolean}
	
	\item \texttt{supercell:} Supercell size and shape to use. This can either be a single int, a list of three integers or a 3x3 matrix of integers. For a single number a diagonal matrix will be generated with the integer repeated on the diagonals. For a list of three numbers a diagonal matrix will be generated where the diagonal elements are set to the list. A matrix will be used directly as is. Default is a 3x3 identity matrix. \deftype{identity}{matrix}
	
	\item \texttt{out\_folder:} Name for the output folder used to store the structural input files generated. \deftype{./muon-airss-out}{string}
	
	\item \texttt{geom\_steps:} Maximum number of geometry optimisation steps.
	\deftype{30}{int} 
	
	\item \texttt{geom\_force\_tol:} Force tolerance for each geometry optimisation in eV/\AA.\deftype{0.05}{float}
	
	\item \texttt{clustering\_method:} Clustering method to use to process results. The options are HIER (for hierarchical clustering) and KMEANS (for k-means clustering). 
	\deftype{hier}{string}
	
	\item \texttt{clustering\_hier\_t:} Normalised $t$ parameter for hierarchical clustering. Higher $t$ will produce a smaller number of bigger clusters. \deftype{0.3}{float}
	
	\item \texttt{clustering\_kmeans\_k:} Expected number of clusters for k-means clustering. \deftype{4}{int}
	
	\item \texttt{clustering\_save\_min}: If True, save the minimum energy structure for each cluster as a separate file. \deftype{false}{boolean}
	
	\item \texttt{clustering\_save\_format}: Extension of file format in which to save the minimum energy structures if \texttt{clustering\_save\_min} is True. \deftype{cif}{string}	
	
	\item \texttt{castep\_command}: Command used to run the CASTEP executable on the system. \deftype{castep.serial}{string}
	
	\item \texttt{dftb\_command}: Command used to run the DFTB+ executable on the system. \deftype{dftb+}{string}
	
	\item \texttt{script\_file}: Path of a submission script template to copy into each individual generated structure's folder, for use with submissions systems on HPC machines. Any literal instance of the string \texttt{\{seedname\}} will be replaced with the name of the structure in that folder, which allows to create submission scripts for batches of CASTEP structures.
	\deftype{NONE}{string}
	
	\item \texttt{castep\_param}: Path to a CASTEP parameter file to use for all calculations.
	\deftype{NONE}{string}
	
	\item \texttt{dftb\_set}: Slater-Koster parametrization to use with DFTB+. Determines which elements can be treated, see \url{dftb.org} for more details. Can currently be \texttt{3ob-3-1} or \texttt{pbc-0-3}.\deftype{3ob-3-1}{string}
	
	\item \texttt{dftb\_optionals}: Additional optional JSON files to activate for the DFTB+ parametrisation. For example, including \texttt{spinpol.json} for \texttt{3ob-3-1} turns spin polarisation on. Should be written like a list of strings (either in square brackets and comma separated, or as a list using a - as bullet and an entry on each line) \deftype{[]}{[str]}
	
	\item \texttt{dftb\_pbc}: Whether to turn on periodic boundary conditions in a DFTB+ calculation. \deftype{true}{boolean}
	
	\item \texttt{k\_points\_grid}: k points grid for periodic system calculations (CASTEP or DFTB+). \deftype{[1,1,1]}{[int]}
	
	\item \texttt{max\_scc\_steps}: Maximum number of self-consistent steps when converging the electronic wavefunction in either CASTEP or DFTB+. \deftype{200}{int}	

\end{itemize}

\section{Supplementary Information}

In this section we will show an example of how a search for the muon stopping sites could be carried out in  $\mathrm{Fe_{3}O_{4}}$. 

\subsection{Run \texttt{pm-symmetry} using $\mathrm{Fe_{3}O_{4}}$'s structural file.}

This would produce an output, whith the symmetry analysis of the special Wyckoff positions in $\mathrm{Fe_{3}O_{4}}$, that is 
shown below:

\begin{lstlisting}[caption=Output of \texttt{pm-symmetry fe3o4.cell}]
Wyckoff points symmetry report for fe3o4.cell
Space Group International Symbol: Fd-3m
Space Group Hall Number: 525
Absolute			Fractional		Hessian constraints
[0. 0. 0.]	[0. 0. 0.]	none
[0.     2.0895 2.0895]	[0.   0.25 0.25]	none
[0.    4.179 4.179]	[0.  0.5 0.5]	none
[0.     6.2685 6.2685]	[0.   0.75 0.75]	none
[1.04475 1.04475 5.22375]	[0.125 0.125 0.625]	isotropic
[1.04475 5.22375 1.04475]	[0.125 0.625 0.125]	isotropic
[2.0895 0.     2.0895]	[0.25 0.   0.25]	none
[2.0895 2.0895 0.    ]	[0.25 0.25 0.  ]	none
[2.0895 4.179  6.2685]	[0.25 0.5  0.75]	none
[2.0895 6.2685 4.179 ]	[0.25 0.75 0.5 ]	none
[3.13425 3.13425 3.13425]	[0.375 0.375 0.375]	isotropic
[3.13425 7.31325 7.31325]	[0.375 0.875 0.875]	isotropic
[4.179 0.    4.179]	[0.5 0.  0.5]	none
[4.179  2.0895 6.2685]	[0.5  0.25 0.75]	none
[4.179 4.179 0.   ]	[0.5 0.5 0. ]	none
[4.179  6.2685 2.0895]	[0.5  0.75 0.25]	none
[5.22375 1.04475 1.04475]	[0.625 0.125 0.125]	isotropic
[5.22375 5.22375 5.22375]	[0.625 0.625 0.625]	isotropic
[6.2685 0.     6.2685]	[0.75 0.   0.75]	none
[6.2685 2.0895 4.179 ]	[0.75 0.25 0.5 ]	none
[6.2685 4.179  2.0895]	[0.75 0.5  0.25]	none
[6.2685 6.2685 0.    ]	[0.75 0.75 0.  ]	none
[7.31325 3.13425 7.31325]	[0.875 0.375 0.875]	isotropic
[7.31325 7.31325 3.13425]	[0.875 0.875 0.375]	isotropic
\end{lstlisting}

As we can see, the list of unoccupied special Wyckoff positions for $\mathrm{Fe_{3}O_{4}}$ is relatively long. Therefore, the symmetry analysis in this case could be useful only if combined with some other piece of known information about the muon stopping site.  For instance, we may know that the stopping site is placed somewhere along a Wyckoff \textit{line} (as it is the case for MnSi).  Otherwise, we would need to test each one of these positions, which might be impractical because, (as it is the case in $\mathrm{Fe_{3}O_{4}}$), the muon stopping site may not be in a Wyckoff position.\\

So, for cases like this one, we need to continue the search for potential muon stopping sites.  The next steps in the procedure include: the generation of muonated structures with muons in random positions; the relaxation of the muon position in each one of this structures and the subsequent performance of a clustering analysis. 

\subsection{Generate Structures with Muons in Random Positions}

We run \texttt{pm-muairss} to generate a set of structure files with muon defects placed in random positions.  This is done by running the line:

\begin{verbatim}
    pm-muairss -t w <fe3o4.cell> <fe3o4.yaml>   
\end{verbatim}

and an example of \texttt{fe3o4.yaml} file for running this calculation is:

\begin{lstlisting}[caption=\texttt{fe3o4.yaml}]
poisson_r: 0.6
name: fe3o4
charged: true
geom_steps: 300
vdw_scale: 0.25
calculator: uep
uep_gw_factor: 4.0
uep_chden: fe3o4.den_fmt
geom_force_tol: 0.05 
clustering_method: hier
clustering_hier_t: 0.2
\end{lstlisting}

The muonated structures generated by this run will be stored in the folders 
\texttt{muon-airss-out/uep/fe3o4\_*},  where * is a numerical label that identifies each particular muonated structure.  The number of structures generated by this procedure will depend on the values of parameters in the \texttt{fe3o4.yaml} file such as \texttt{vdw\_scale} and \texttt{poisson\_r}.\\

The next step is to relax each one of these newly generated muonated structures. 

\subsection{Relaxing the Muon Position in each Structure with \texttt{pm-uep-opt}}

In each one of these newly created \texttt{fe3o4\_*} folders there will be a new \texttt{fe3o4\_*.yaml} file, which will contain instructions for relaxing the muon positions in each one of the muonated structures by running the library \texttt{pm-uep-opt}. Below there is an example of this \texttt{fe3o4\_*.yaml} file:

\begin{lstlisting}[caption=\texttt{fe3o4\_*.yaml}]
chden_path: path-to-folder
chden_seed: fe3o4
geom_steps: 300
gw_factor: 6.0
mu_pos:
- 6.6017194506272086
- 3.924221192037714
- 3.8231862749935717
opt_method: trust-exact
opt_tol: 0.05
particle_mass: 1.67382335232e-27
save_pickle: true
\end{lstlisting}

and we relax each one of these structures by running:

\begin{verbatim}
    pm-uep-opt <fe3o4_*.yaml>   
\end{verbatim}

for each \texttt{fe3o4\_*.yaml} in each one of the newly generated folders\footnote{If the calculation is being run in Linux, the relaxations could be done, for instance, using a Bash script.}.  The result will be output in a \texttt{fe3o4\_*.uep} file such as: 

\newpage

\begin{lstlisting}[caption=\texttt{fe3o4\_*.yaml}]

*********************************
|   UU  UU   EEEEEE    PPPP     |
|   UU  UU   EE        PP  PP   |
|   UU  UU   EEEEEE    PPPP     |
|   UU  UU   EE        PP       |
|    UUUU    EEEEEE    PP       |
*********************************

Unperturbed Electrostatic Potential
optimiser for mu+ stopping site finding

by Simone Sturniolo (2018)

Calculations started on 2019-12-05 15:24:10.820543

Charge distribution loaded from ~/Calculations/UEP_Paper/Fe3O4/fe3o4
Gaussian width factor used: 6.0
Particle mass: 1.67382335232e-27 kg

---------

Performing optimisation with method trust-exact
Tolerance required for convergence: 0.05 eV
Maximum number of steps: 300
Defect starting position: 6.6017194506272086 3.924221192037714 3.8231862749935717 Ang

---------

Optimisation stopped after 8 steps

Final coordinates: 5.789295137667838 4.670038556634775 2.983741171022327 Ang
Final fractional coordinates: 0.6926651277420242 0.5587507246512053 0.3569922434819726
Classical energy: -11.47475742543093 eV
Zero-point energy: 0.17842078166219383 eV
Quantum total energy: -11.296336643768736 eV

Calculation time: 64.172284 s

\end{lstlisting}

These relaxations may take some time.  If the system is small and simple, the relaxations will be fast.  However, if the system is large and sophisticated and there is a large number of structures, these relaxations could be relatively expensive to run. 

In any case, once the relaxations are ready, we need to perform the clustering analysis. This is done by running the line 

\begin{verbatim}
    pm-muairss -t r <fe3o4.cell> <fe3o4.yaml>   
\end{verbatim}

from the folder where all the structures were generated\footnote{In our case, this is the folder from where we can see the \texttt{muon-airss-out} folder.}. This generates two new files: 

\begin{itemize}
\item \texttt{fe3o4\_clusters.text}: This file contains the structures that form each of predicted of the clusters that are associated to potential muon stopping sites.  This file  also has information about which would be a representative structural file for each stopping site.  

\item \texttt{fe3o4\_fe3o4\_uep\_clusters.dat}: This file has the average energy, minimum structure energy and standard deviation of the energy for each of the clusters.
\end{itemize}

Examples of this two files are below:

\begin{lstlisting}[caption=\texttt{fe3o4\_clusters.text}]

****************************
|                          |
|         MUAIRSS          |
|    Clustering report     |
|                          |
****************************

Name: fe3o4
Date: 2019-12-05 18:02:33.530251
Structure file(s): fe3o4-out.cell
Parameter file: fe3o4.yaml

Clustering method: Hierarchical
    t = 0.2


*******************

Clusters for fe3o4:
CALCULATOR: uep
	2 clusters found


	-----------
	Cluster 1
	-----------
	Structures: 13

	Energy (eV):
	Minimum		Average		StDev
	-9.90		-9.90		0.00

	Minimum energy structure: fe3o4_29


	Structure list:
	fe3o4_2	fe3o4_13	fe3o4_14	fe3o4_29	
	fe3o4_38	fe3o4_57	fe3o4_73	fe3o4_88	
	fe3o4_115	fe3o4_120	fe3o4_132	fe3o4_137	
	fe3o4_145	

	-----------
	Cluster 2
	-----------
	Structures: 139

	Energy (eV):
	Minimum		Average		StDev
	-11.47		-11.47		0.00

	Minimum energy structure: fe3o4_48


	Structure list:
	fe3o4_1	fe3o4_3	fe3o4_4	fe3o4_5	
	fe3o4_6	fe3o4_7	fe3o4_8	fe3o4_9	
	fe3o4_10	fe3o4_11	fe3o4_12	fe3o4_15	
	fe3o4_16	fe3o4_17	fe3o4_18	fe3o4_19	
	fe3o4_20	fe3o4_21	fe3o4_22	fe3o4_23	
	fe3o4_24	fe3o4_25	fe3o4_26	fe3o4_27	
	fe3o4_28	fe3o4_30	fe3o4_31	fe3o4_32	
	fe3o4_33	fe3o4_34	fe3o4_35	fe3o4_36	
	fe3o4_37	fe3o4_39	fe3o4_40	fe3o4_41	
	fe3o4_42	fe3o4_43	fe3o4_44	fe3o4_45	
	fe3o4_46	fe3o4_47	fe3o4_48	fe3o4_49	
	fe3o4_50	fe3o4_51	fe3o4_52	fe3o4_53	
	fe3o4_54	fe3o4_55	fe3o4_56	fe3o4_58	
	fe3o4_59	fe3o4_60	fe3o4_61	fe3o4_62	
	fe3o4_63	fe3o4_64	fe3o4_65	fe3o4_66	
	fe3o4_67	fe3o4_68	fe3o4_69	fe3o4_70	
	fe3o4_71	fe3o4_72	fe3o4_74	fe3o4_75	
	fe3o4_76	fe3o4_77	fe3o4_78	fe3o4_79	
	fe3o4_80	fe3o4_81	fe3o4_82	fe3o4_83	
	fe3o4_84	fe3o4_85	fe3o4_86	fe3o4_87	
	fe3o4_89	fe3o4_90	fe3o4_91	fe3o4_92	
	fe3o4_93	fe3o4_94	fe3o4_95	fe3o4_96	
	fe3o4_97	fe3o4_98	fe3o4_99	fe3o4_100	
	fe3o4_101	fe3o4_102	fe3o4_103	fe3o4_104	
	fe3o4_105	fe3o4_106	fe3o4_107	fe3o4_108	
	fe3o4_109	fe3o4_110	fe3o4_111	fe3o4_112	
	fe3o4_113	fe3o4_114	fe3o4_116	fe3o4_117	
	fe3o4_118	fe3o4_119	fe3o4_121	fe3o4_122	
	fe3o4_123	fe3o4_124	fe3o4_125	fe3o4_126	
	fe3o4_127	fe3o4_128	fe3o4_129	fe3o4_130	
	fe3o4_131	fe3o4_133	fe3o4_134	fe3o4_135	
	fe3o4_136	fe3o4_138	fe3o4_139	fe3o4_140	
	fe3o4_141	fe3o4_142	fe3o4_143	fe3o4_144	
	fe3o4_146	fe3o4_147	fe3o4_148	fe3o4_149	
	fe3o4_150	fe3o4_151	fe3o4_152	

	----------

	Similarity (ranked):
	0 <--> 1 (distance = 1.602)

--------------------------


==========================
\end{lstlisting}

\begin{lstlisting}[caption=\texttt{fe3o4\_fe3o4\_uep\_clusters.dat}]
1       13      -9.897753370173747      -9.897721148028158      5.260352456001172e-05
2       139     -11.474801196064512     -11.474700683287459     0.00015287319327130454
\end{lstlisting}

\bibliographystyle{abbrv}
\bibliography{aipsamp}

\end{document}